\documentclass[manuscript]{aastex}
\usepackage{amssymb}

\begin{document}
\title{PROPERTIES AND KEPLERIAN ROTATION OF THE HOT CORE IRAS 20126+4104}
\author{Jin-Long Xu\altaffilmark{1,2,3}, Jun-Jie Wang\altaffilmark{1,2} and Chang-Chun Ning \altaffilmark{2,4}}
\altaffiltext{1}{National Astronomical Observatories, Chinese
Academy of Sciences, Beijing 100012, China} \altaffiltext{2}{NAOC-TU
Joint Center for Astrophysics, Lhasa,  Tibet, 850000, China}
\altaffiltext{3}{Graduate University of the Chinese Academy of
Sciences, Beijing, 100080, China}\altaffiltext{4}{Tibet University,
Lhasa,  Tibet, 850000, China} \email{xujl@bao.ac.cn}

\begin{abstract} We present Submillimeter Array observations
of the massive star-forming region IRAS 20126+4104 in the millimeter
continuum and in several molecular line transitions. With the SMA
data, we have detected nine molecular transitions, including DCN,
$\rm CH_{3}OH$, $\rm H_{2}CO$, and $\rm HC_{3}N$ molecules, and
imaged each molecular line. From the 1.3 mm continuum emission a
compact millimeter source is revealed, which is also associated with
$\rm H_{2}O$, OH, and $\rm CH_{3}OH$ masers. Using a rotation
temperature diagram (RTD), we derive that the rotational temperature
and the column density of $\rm CH_{3}OH$ are 200 K and 3.7 $\times$
$10^{17}$ $\rm cm^{-2}$, respectively. The calculated results and
analysis further indicate that a hot core coincides with IRAS
20126+4104. The position-velocity diagrams of $\rm H_{2}CO$
($3_{_{0,3}}$-$2_{_{0,2}}$) and $\rm HC_{3}N$ ($25$-$24$) clearly
present Keplerian rotation. Moreover, $\rm H_{2}CO$
($3_{_{0,3}}$-$2_{_{0,2}}$) is found to trace the disk rotation for
the first time.

\end{abstract}

\section{Introduction}
Massive star formation, especially in the earliest phase, is poorly
understood, due to the large distances involved ($\geq1$ kpc),
complex cluster formation environments and shorter evolutionary
timescales of massive stars.  Hot cores represent the early phase in
the process of massive star-formation (Qin et al. 2010 ). The hot
cores are defined as compact ( $\leq$ 0.1 pc, n $\geq$ $10^{7}$ $\rm
cm^{-3}$), relatively high temperature ($T\rm_{k}$ $\geq$ 100 K)
cloud cores (Kurtz et al. 2000). Thus, observation of hot cores with
high angular resolution and various molecular lines can provide
important information on the early evolutionary phase of massive
star formation.

IRAS 20126+4104, at a distance of 1.7 kpc from the Sun (Dame \&
Thaddeus, 1985; Wilking et al. 1989; Moscadelli et al. 2011), is a
massive star-forming region with a bolometric luminosity of
1.3$\times$10$^{4}L_{\odot}$. Previous observations of $\rm NH_{3}$,
$\rm CH_{3}CN$ and CS molecular lines toward IRAS 20126+4104
indicated that a Keplerian disk was detected around a $\sim$ 7-24
$M_{\odot}$ point mass (Zhang et al. 1998; Keto \& Zhang 2010;
Cesaroni et al. 1999, 2005). Traced in CO, $\rm NH_{3}$, SiO, $\rm
HCO^{+}$, $\rm CH_{3}OH$ and $\rm H_{2}$ molecular lines (Shinnaga,
et al. 2008; Su et al. 2007; Zhang et al. 1999; Cesaroni et al.
1997, 1999, 2005; Shepherd et al. 2000), a bipolar jet/outflow
system was widely investigated toward IRAS 20126+4104. In addition,
$\rm H_{2}O$, OH and $\rm CH_{3}OH$ masers in IRAS 20126+4104 were
revealed from observations at centimeter bands (Tofani et al. 1995;
Trinidad et al. 2005; Minier et al. 2001; Edris et al. 2005). The
$\rm CH_{3}OH$ masers might be associated with a hot core (Minier et
al. 2001). From VLA observations toward IRAS 20126+4104, Tofani et
al. (1995) did not detect the 1.3 and 3.6 cm continuum emission, but
Hofner et al. (1999) detected two elongated structures at the 3.6 cm
continuum band. They concluded that the emission was associated with
thermal ionized jets and not from a UC$\rm H \scriptstyle II$
region. Moreover, IRAS 20126+4104 has the typical colors of UC$\rm H
\scriptstyle II$ (Wood \& Churchell 1989), suggesting that IRAS
20126+4104 may be a precursor of a UC$\rm H \scriptstyle II$ region
(Molinari et al. 1996). Thus, IRAS 20126+4104 provides us with an
opportunity to study the early evolution of massive star formation.

We have carried out multiline observations toward the massive
star-forming region IRAS 20126+4104 with the Submillimeter Array.
Various molecular lines are used to investigate the physical and
chemical processes of IRAS 20126+4104. So far, only a few molecular
lines exclusively indicate rotational motions (Beuther  2007). In
this paper, we find new molecular line ( $\rm H_{2}CO$ ) which can
trace the Keplerian disk. In Sect. 2, we summarize the observations.
In Sect. 3, we give the general results. We discuss core properties
and core kinematics in Sect. 4.1 and Sect. 4.2, respectively. In
Sect. 5, we summarize our main conclusions.

\section{OBSERVATIONS AND DATA REDUCTION}

The data are from SMA archive. Observations toward IRAS 20126+4104
were carried out with the SMA on 2004 August 23, at 218 (lower
sideband) and 228 GHz (upper sideband). The typical system
temperature was 130 K. The compact configuration with seven antennas
was used. The phase center was R.A(J2000.0)= $\rm 20
^{h}14^{m}26^{s}.00$ and decl.(J2000.0) = $\rm 41
^{\circ}13^{\prime}31^{\prime\prime}.50$. The spectral resolution is
0.8125 MHz, corresponding to a velocity resolution of 1.1 km $\rm
s^{-1}$. The absolute flux density scales were determined from
observations of Uranus, and the bright quasar 3C 279 was used for
bandpass calibration. QSO 2015+371 and QSO 2202+422 were observed
for the antenna gain corrections. The calibration and imaging were
performed in Miriad. The continuum map was made using the line-free
channels. The spectral cubes were constructed using the
continuum-subtracted spectral channels. Self-calibration was
performed to the continuum data. The gain solutions from the
continuum were applied to the line data. The synthesized beam size
of  continuum and line images with robust weighting was
approximately $1^{\prime\prime}.26$ $\times$ $0^{\prime\prime}.96$
(P.A. = $89^{\circ}.0$).

\section{RESULTS}
\subsection{Continuum Emission at 1.3 mm}
Figure 1  is the 1.3 mm continuum map of IRAS 20126+4104 obtained by
combining the lower and upper sideband data. The continuum emission
shows a compact source, which is unresolved in the beam at 1.3 mm.
By using a two-dimensional Gaussian fit for the continuum emission,
we obtained that the total flux density is 0.62 $\pm$ 0.03 Jy, the
deconvolved source size is 0.8$^{\prime\prime}$ $\times$
0.5$^{\prime\prime}$ (P.A. = 87.9$^{\circ}$), and the peak position
is R.A.(J2000) = 20$\rm ^{h}$14$\rm ^{m}$26.$\rm ^{s}$040
($\Delta$R.A. = $\pm$0.04$^{\prime\prime}$), decl.(J2000) =
+41$^{\circ}$13$^{\prime}$32.$^{\prime\prime}$55 ($\Delta$decl. =
$\pm$0.03$^{\prime\prime}$) with an intensity of 0.46 $\pm$ 0.05 Jy
beam$^{-1}$.  The peak position of the 3.3 mm continuum emission is
R.A.(J2000) = 20$\rm ^{h}$14$\rm ^{m}$26.$\rm ^{s}$008 ($\Delta$R.A.
= $\pm$0.01$^{\prime\prime}$), decl.(J2000) =
+41$^{\circ}$13$^{\prime}$32.$^{\prime\prime}$73 ($\Delta$decl. =
$\pm$0.01$^{\prime\prime}$) (Cesaroni et al. 1997). For the 7 mm
continuum emission (Hofner et al. 1999), the peak position  is
R.A.(J2000) = 20$\rm ^{h}$14$\rm ^{m}$26.$\rm ^{s}$026, decl.(J2000)
= +41$^{\circ}$13$^{\prime}$32.$^{\prime\prime}$70. The three
positions coincide within the uncertainties. $\rm H_{2}O$, OH and
$\rm CH_{3}OH$ masers in IRAS 20126+4104 have been detected from
observations at centimeter wavelengths (Tofani et al. 1995; Trinidad
et al. 2005; Minier et al. 2001; Edris et al. 2005). In Figure 1,
$\rm H_{2}O$ masers are shown with the blue filled triangles. The
green open squares present OH masers. The filled black circles show
$\rm CH_{3}OH$ masers. Most of the masers are associated with the
1.3 mm continuum emission, but are not located at the peak position
of the 1.3 continuum emission.

\subsection{Molecular Line Emission}
In the frequency coverage of the 2 GHz band, deuterated hydrogen
cyanide ($\rm DCN$), formaldehyde ($\rm H_{2}CO$), cyanoacetylene
($\rm HC_{3}N$) and methanol ($\rm CH_{3}OH$) emission lines were
identified by use of the CDMS and JPL catalog ( M\"{u}ller et al.
2005; Pickett et al. 1998), as done by other authors in the same
frequency range toward Orion (Sutton et al. 1985) and Sgr B2 (Qin et
al. 2008). The transition, rest frequency and upper level energy
($E_{\rm u}$) of each molecular line are presented in Table 1.

Molecular line images can provide valuable information on the
spatial distribution of molecules and kinematics, then we imaged the
molecular lines of  DCN, $\rm CH_{3}OH$, $\rm H_{2}CO$, and $\rm
HC_{3}N$. The 1.3 mm continuum emission has been subtracted before
computing these maps. Figure 2 presents the integrated intensity
maps of each molecular line. $``\times"$ indicates the peak position
of the continuum emission. In Figure 2, each line emission is
spatially coincident with the continuum emission.

Spectra of molecular emission are present in Figure 3. The spectra
were extracted from the peak positions of each integrated intensity
map. We have made Gaussian fits to all the spectra. The central line
velocity ($V_{\rm LSR}$), peak intensity ($I_{\rm p}$), and full
width at half-maximum ($\Delta V$) are summarized in Table 1.

\section{DISCUSSION}
\subsection{Core Properties}
Our interferometric continuum and multiline observations toward the
massive star-forming region IRAS 20126+4104 reveal a compact
millimeter source, which is associated with $\rm H_{2}O$, OH, and
$\rm CH_{3}OH$ masers. Millimeter continuum emission mainly traces
warm dust heated by the embedded young stellar object. Four
transitions of $\rm CH_{3}OH$ have been detected in IRAS 20126+4104,
so we can use a rotation temperature diagram (RTD) to estimate the
rotation temperature and the column density. Assuming local
thermodynamic equilibrium (LTE), lines being optically thin and gas
emission filling the beam, the rotation temperature and the
beam-averaged column density can be expressed by (Goldsmith $\&$
Langer 1999; Liu et al. 2002; Qin et al. 2010)
\begin{equation} \mathit{\rm ln(\it \frac{N_{ u}}{g_{u}})}=\rm ln(\it \frac{N_{T}}{Q_{rot}})-\frac{E_{ u}}{T_{ rot}},
\end{equation}
Where $N_{u}$ is the column density of the upper energy level,
$g_{u}$ is the degeneracy factor in the upper energy level, $N_{T}$
is the total beam-averaged column density, $Q_{rot}$ is the
rotational partition function, $E_{u}$ is the upper level energy in
K, and $T_{rot}$ is the rotation temperature. Figure 4 is the
rotation temperature diagram. A linear least-squares fit is
performed toward the four transitions of $\rm CH_{3}OH$. The RTD can
be corrected by multiplying $N_{ u}/g_{u}$ by the optical depth
correction factor $C_{\tau}=\tau/(1-e^{\tau})$, where $\tau$ is the
optical depths. According to Qin et el. (2010), we iteratively
applied the $C_{\tau}$ correction to the RTD until the solution
converged. After the optical depth is corrected, the rotation
temperature and the beam-averaged column density are 200 $\pm$ 57 K
and (3.7 $\pm$ 1.3) $\times$ $10^{17}$ $\rm cm^{-2}$, respectively.
The  rotation temperature of  200 $\pm$ 57 K is the same as that
estimated in the $\rm CH_{3}OH$ (2-1) and (5-4) lines (171 K;
Cesaroni et al. 2005), and in the $\rm CH_{3}CN$ lines (201 K;
Cesaroni et al. 1997).

In the Orion molecular cloud, the strongest $\rm CH_{3}OH$ emission
is from the compact ridge (Schilke et al. 1997), giving fractional
abundances relative to $\rm H_{2}$ of 1.2 $\times$ $10^{-7}$. We
adopt 0.007 pc as the deconvolved size of the $\rm CH_{3}OH$
emission derived from the integrated intensity maps of $\rm
CH_{3}OH$, which is smaller than the canonical size of 0.1 pc that
is commonly quoted for hot molecular cores (Kurtz et al. 2000). By
using the fractional abundance in the Orion molecular cloud, the
calculated $\rm H_{2}$ density is 4 $\times10^{7}$ $\rm cm^{-3}$.
The relatively higher gas temperature (200 $\pm$ 57 K), $\rm H_{2}$
density ($10^{7} \rm cm^{-3}$), and smaller size (0.007 pc) indicate
a hot core in this region. Following Gibb et al. (2000), we adopt 3
$\times$ $10^{24}$ $\rm cm^{-2}$ as a typical $\rm H_{2}$ column
density. We obtain that the fractional abundance of $\rm CH_{3}OH$
relative to $\rm H_{2}$ is 8 $\times$ $10^{-6}$. The fractional
abundance of $\rm CH_{3}OH$ is close to that in the Orion hot core
(Blake et al. 1987), but higher than that in the Sgr B2 core (Qin et
al. 2008).

\subsection{Core Kinematics}

The different characteristics of molecular lines may be used as
tools to trace various physical processes of massive star formation.
To understand the kinematic signatures of the rotating disk, we use
the $\rm H_{2}CO$ ($3_{_{0,3}}$-$2_{_{0,2}}$) and $\rm HC_{3}N$
($25$-$24$) lines to construct a position-velocity diagram cut
through the continuum peak position at a position angle of
$58^{\circ}$.  In Figure 5, the continuum emission is redshifted to
the southwest and blueshifted to the northeast. The
position-velocity diagrams of $\rm H_{2}CO$
($3_{_{0,3}}$-$2_{_{0,2}}$) and $\rm HC_{3}N$ ($25$-$24$) clearly
present Keplerian rotation. The Keplerian rotation traced in the
$\rm HC_{3}N$ ($25$-$24$) line is extremely similar to the model
disk constructed by Zhang et al. (1998). So far, for many sources
only a few molecular lines exclusively indicated rotating motions in
massive star formation (Beuther 2007), e.g., $\rm CH_{3}CN$, $\rm
C^{34}S$, $\rm NH_{3}$, $\rm HCOOCH_{3}$, $\rm C^{17}O$, $\rm
H_{2}^{18}O$, $\rm HN_{13}C$, $\rm HC_{3}N$ and $\rm CH_{3}OH$. Here
we find a new molecule ($\rm H_{2}CO$ $3_{_{0,3}}$-$2_{_{0,2}}$),
which can be used to investigate the rotational signatures of a
disk.

In Figure 5, a velocity gradient in the NE-SW direction is clearly
depicted. The velocities of the two emission peaks are -6.0 and -2.0
$\rm km$ $\rm s^{-1}$ with $0.60^{\prime\prime}$ separation for the
$\rm H_{2}CO$ ($3_{_{0,3}}$-$2_{_{0,2}}$) molecule, and -5.6 and
-3.0 $\rm km$ $\rm s^{-1}$ with $0.55^{\prime\prime}$ separation for
the $\rm HC_{3}N$ ($25$-$24$) molecule, respectively. Assuming
equilibrium between the rotational and gravitational force at the
outer radius of the Keplerian disk:
\begin{equation} \mathit{M_{\rm rot}}=\frac{V^{2}r}{G}=2.3 \times 10^{2}\ V^{2}(\rm km \: s^{-1})\it r(\rm pc)\it \:M_{\odot},
\end{equation}
where $V$ is the velocity difference, $r$ is the spatial separation
of the emission peaks and $G$ is the gravitational constant. The
derived dynamical mass is 18.0 $M_{\odot}$ and 7.1 $M_{\odot}$ in
the $\rm H_{2}CO$ ($3_{_{0,3}}$-$2_{_{0,2}}$) and $\rm HC_{3}N$
($25$-$24$) molecular lines, respectively. The calculated dynamical
mass is smaller than a mass of $\sim$ 20 $M_{\odot}$ obtained by
Zhang et al (1998) in the $\rm NH_{3}$ molecular lines.

\section{SUMMARY}
We have observed the massive star-forming region IRAS 20126+4104 in
the millimeter continuum and in several molecular line transitions.
The 1.3 mm continuum emission reveals a compact millimeter source
which is associated with $\rm H_{2}O$, OH, and $\rm CH_{3}OH$
masers. From the SMA data, nine molecular transitions, including
DCN, $\rm CH_{3}OH$, $\rm H_{2}CO$, and $\rm HC_{3}N$ molecules are
detected. The rotational temperature and the column density of $\rm
CH_{3}OH$ emission are 200 K and 3.7 $\times$ $10^{17}$ $\rm
cm^{-2}$, respectively. The calculated results and analysis further
indicate that a hot core coincides with IRAS 20126+4104. A Keplerian
rotational disk in the IRAS 20126+4104 hot core is verified in the
$\rm H_{2}CO$ ($3_{_{0,3}}$-$2_{_{0,2}}$) and $\rm HC_{3}N$
($25$-$24$) lines. $\rm H_{2}CO$ ($3_{_{0,3}}$-$2_{_{0,2}}$) is
found to trace the disk rotation for the first time. The dynamical
mass of the Keplerian rotational disk is 18.0 $M_{\odot}$ and 7.1
$M_{\odot}$, determined in the $\rm H_{2}CO$
($3_{_{0,3}}$-$2_{_{0,2}}$) and $\rm HC_{3}N$ ($25$-$24$) molecular
lines, respectively.

\acknowledgments We thank the SMA staff for the observations and the
anonymous referee whose insights greatly improved this manuscript.
We also thank Dr. H. Shi and other people for providing help with
the data reduction and analysis. This work was supported by the
National Natural Science Foundation of China under Grant
No.10473014.

\clearpage
\begin{figure}[]
\vspace{0mm}\includegraphics[angle=270, scale=.4]{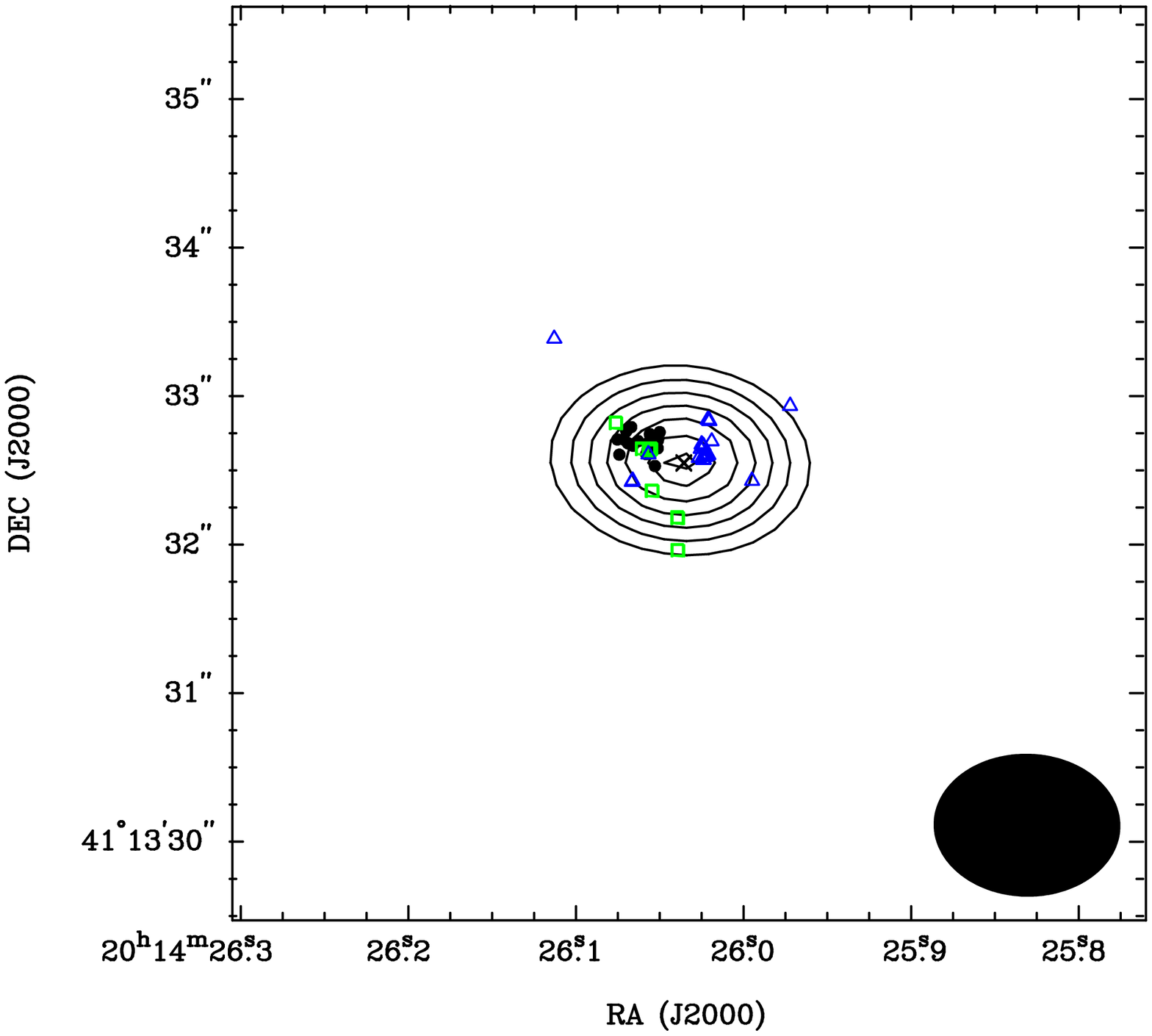}
\caption{Continuum map toward IRAS 20126+4104 at 1.3mm obtained by
combining the data from both sidebands (218 and 228 GHz). The
contours are at -3, 3, 4, 5, 6, 7, 8, and 9 $\sigma$. The rms noise
level is 0.05 $\rm Jy\ beam^{-1}$($1\sigma$). The synthesized beam
is $1.3^{\prime \prime}\times0.8^{\prime \prime}$, P.A.=
88.9$^{\circ}$ ($lower$ $right$ $corner$).  $``\times"$ indicates
the peak position of the continuum  from  the Gaussian fit. $\rm
H_{2}O$ masers are shown with blue filled triangles. The green open
squares present OH masers. The filled black circles show $\rm
CH_{3}OH$ masers.}
 \vspace{3mm}
\end{figure}

\begin{figure}[h]
\includegraphics[angle=270,scale=.8]{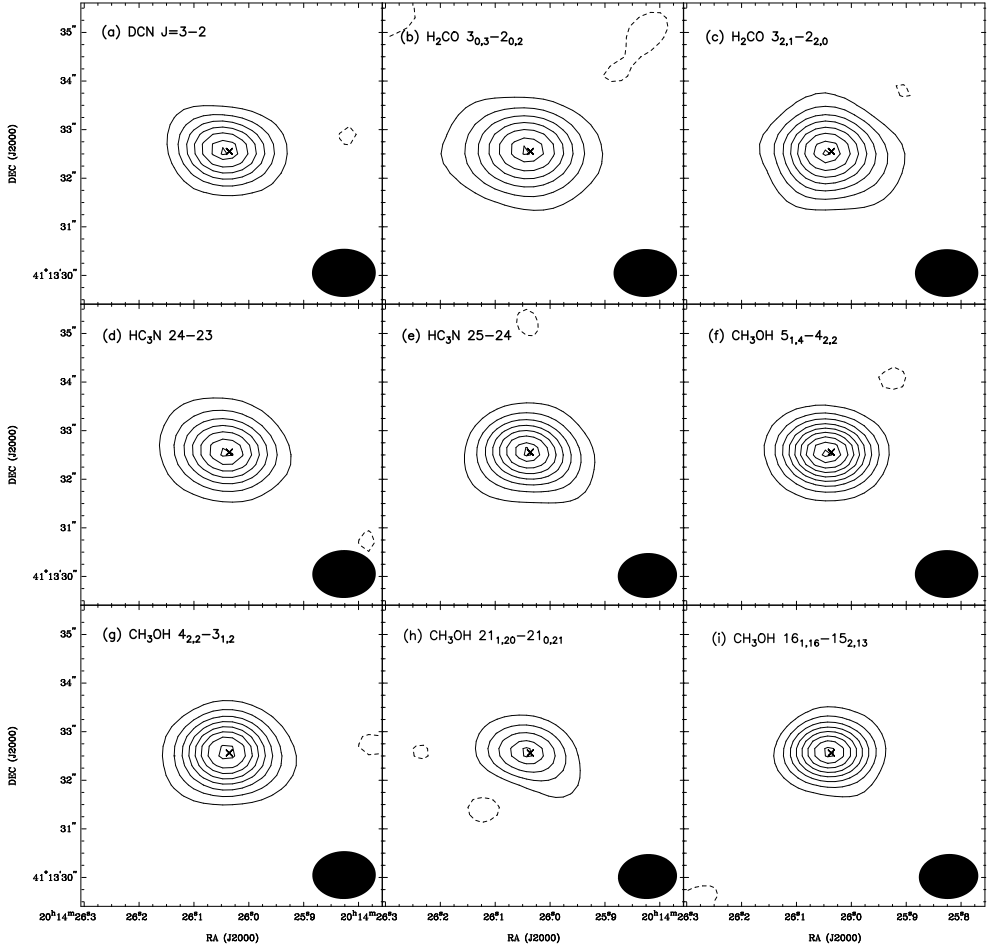}
\vspace{-35mm}\caption{(a) The DCN 3-2 integrated intensity map. The
contour levels are at -2.2, 2.2, 4.4, 6.6, 8.8, 11.0, 13.2 and 14.3
Jy $\rm beam ^{-1}$ km $\rm s^{-1}$. (b) The $\rm H_{2}CO$
$3_{_{0,3}}$-$2_{_{0,2}}$ integrated intensity map. The contour
levels are at -1.0, 2.5, 5.0, 7.5, 9.0, 12.5, 15.0 and 16.2 Jy $\rm
beam ^{-1}$ km $\rm s^{-1}$. (c) The $\rm H_{2}CO$
$3_{_{2,1}}$-$2_{_{2,0}}$ integrated intensity map. The contour
levels are at -1.5, 2.4, 4.8, 7.2, 9.6, 12.0, 14.4, 16.8 and 18.1 Jy
$\rm beam ^{-1}$ km $\rm s^{-1}$. (d) The $\rm HC_{3}N$ $24$-$23$
integrated intensity map. The contour levels are at -1.5, 2.3, 4.5,
6.7, 8.9, 11.1, 13.3 and 15.1 Jy $\rm beam ^{-1}$ km $\rm s^{-1}$.
(e) The $\rm HC_{3}N$ $25$-$24$ integrated intensity map. The
contour levels are at -1.6, 2.1, 4.2, 6.3, 8.4, 10.5, 12.6, 14.7 and
15.9 Jy $\rm beam ^{-1}$ km $\rm s^{-1}$. (f) The $\rm CH_{3}OH$
$5_{_{1,4}}$-$4_{_{2,2}}$ integrated intensity map. The contour
levels are at -1.1, 2.2, 4.4, 6.6, 8.8, 11.0, 13.2, 15.4, 17.6, and
19.3 Jy $\rm beam ^{-1}$ km $\rm s^{-1}$. (g) The $\rm CH_{3}OH$
$5_{_{1,4}}$-$4_{_{2,2}}$ integrated intensity map. The contour
levels are at -2.1, 2.8, 5.6, 8.4, 11.2, 14.0, 16.8, 19.6 and 22.8
Jy $\rm beam ^{-1}$ km $\rm s^{-1}$. (h) The $\rm CH_{3}OH$
$21_{_{1,20}}$-$21_{_{0,21}}$ integrated intensity map. The contour
levels are at -1.4, 1.8, 3.6, 5.4, 7.2,  and 8.2 Jy $\rm beam ^{-1}$
km $\rm s^{-1}$. (i) The $\rm CH_{3}OH$
$16_{_{1,16}}$-$15_{_{2,13}}$ integrated intensity map. The contour
levels are at -1.5, 2.0, 4.1, 6.1, 8.2, 10.2, 12.2, 14.3 and 16.1 Jy
$\rm beam ^{-1}$ km $\rm s^{-1}$. $``\times"$ indicates the position
of the continuum peak. The synthesized beam is shown in the lower
right corner of each map.}
\end{figure}

\begin{table*}[]
\begin{center}
\tabcolsep 1mm\caption{Observed Parameters of Each Line}
\begin{tabular}{lcccccccc}
\tableline\tableline
Molecule   & Transition      & Rest Frequency   & $E_{\rm u}$  & $V_{\rm_{LSR}}$  & $I_{\rm p}$    & $\Delta V$ & Channel rms \\
         &           &(GHz)     &{\rm (K)}    &${\rm (km\ s^{-1})}$ &($\rm Jy\ beam^{-1})$& (${\rm km\ s^{-1})}$ & (Jy beam$^{-1}$) \\
  \hline\noalign{\smallskip}
$\rm DCN$       & $3-2$                        & 217.239   & 20.9 & -3.4(0.2)   & 1.65(0.08) & 8.0(0.4) & 0.2\\  
$\rm H_{2}CO$   & $3_{_{0,3}}$-$2_{_{0,2}}$    & 218.222   & 21.0 & -3.8(0.3)   & 2.38(0.19) & 8.6(0.8) & 0.3\\  
                &$3_{_{2,1}}$-$2_{_{2,0}}$     & 218.760   & 68.1 & -5.1(0.3)   & 2.09(0.13) & 7.5(0.6) & 0.2\\  
$\rm HC_{3}N$   & $24$-$23$                    & 218.325   & 131.0& -4.1(0.2)   & 1.78(0.11) & 7.8(0.6) & 0.2\\  
                & $25$-$24$                    & 227.419   & 141.9& -4.6(0.1)   & 2.53(0.09) & 6.7(0.3) & 0.2\\  
$\rm CH_{3}OH$  &$5_{_{1,4}}$-$4_{_{2,2}}$     & 216.946   & 55.9 & -3.8(0.2)   & 2.02(0.12) & 8.2(0.5) & 0.2\\  
                & $4_{_{2,2}}$-$3_{_{1,2}}$    & 218.440   & 45.5 & -4.1(0.1)   & 3.01(0.17) & 7.3(0.5) & 0.2\\  
                & $21_{_{1,20}}$-$21_{_{0,21}}$& 227.095   & 557.1& -3.7(0.4)   & 0.87(0.06) & 9.3(0.9)& 0.1\\  
                &$16_{_{1,16}}$-$15_{_{2,13}}$ & 227.815   & 327.2& -4.0(0.2)   & 1.62(0.07) & 8.2(0.4) & 0.1\\  
\tableline
\end{tabular}\end{center}
\vspace{10mm}
\end{table*}

\begin{figure*}[]
\includegraphics[angle=180,scale=.78]{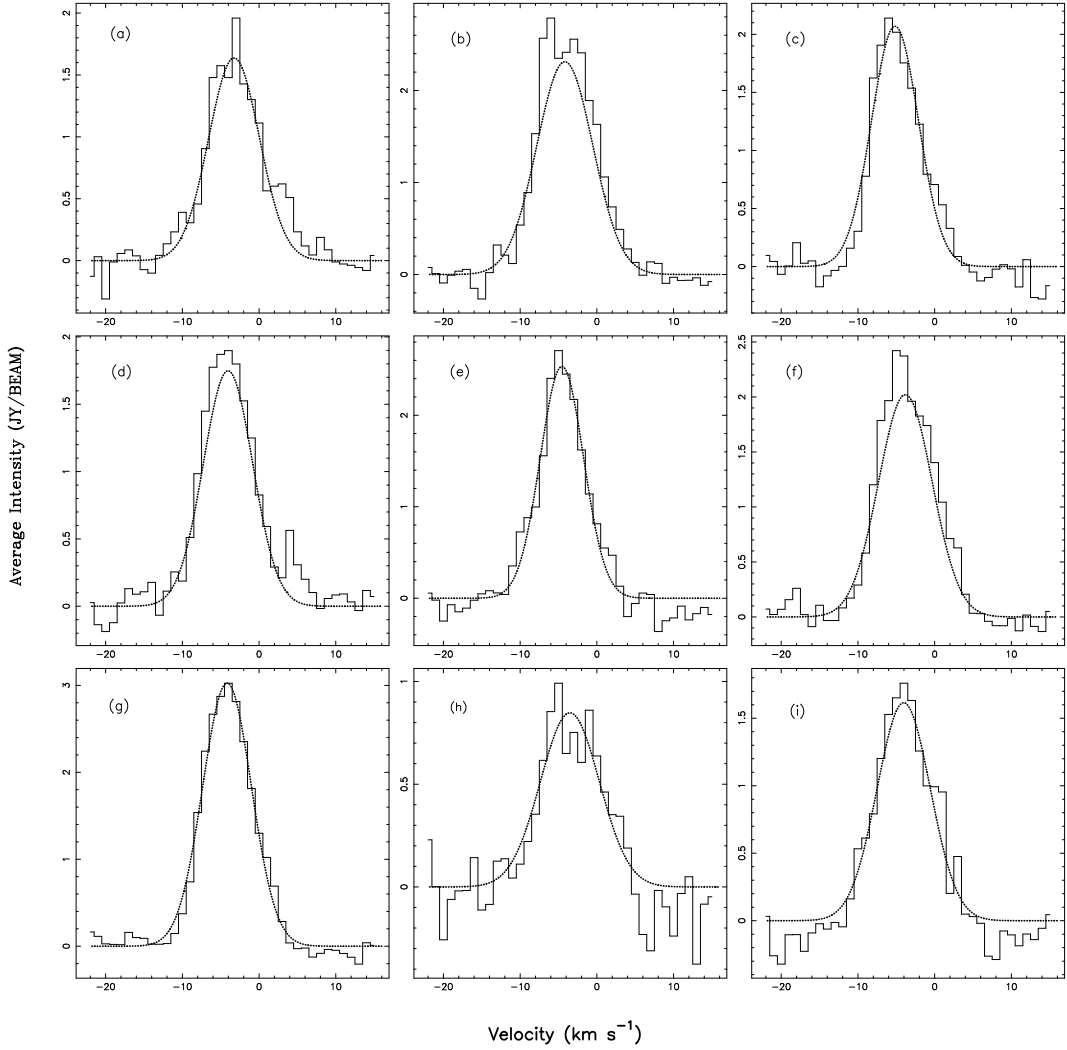}
\vspace{-60mm}\caption{Molecular spectra at the peak position of
different molecular core. The solid and dashed curves are the
observed spectra and the Gaussian fitting to the spectra,
respectively. (a) DCN 3-2. (b) $\rm H_{2}CO$
$3_{_{0,3}}$-$2_{_{0,2}}$. (c) $\rm H_{2}CO$
$3_{_{2,1}}$-$2_{_{2,0}}$. (d) $\rm HC_{3}N$ $24$-$23$. (e) $\rm
HC_{3}N$ $25$-$24$. (f) $\rm CH_{3}OH$ $5_{_{1,4}}$-$4_{_{2,2}}$.
(g) $\rm CH_{3}OH$ $4_{_{2,2}}$-$3_{_{1,2}}$. (h) $\rm CH_{3}OH$
$21_{_{1,20}}$-$21_{_{0,21}}$. (i) $\rm CH_{3}OH$
$16_{_{1,16}}$-$15_{_{2,13}}$.}
\end{figure*}

\begin{figure}[]
\vspace{-6mm}
\includegraphics[angle=0,scale=.5]{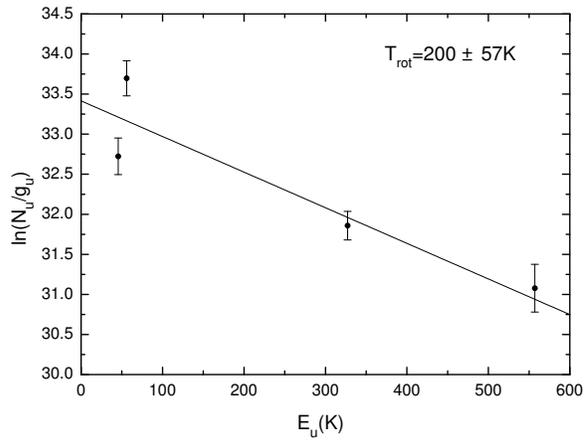}
\vspace{-6mm}\caption{ Population temperature diagram of the
observed $\rm CH_{3}OH$ transitions. The vertical bars mark the $\rm
ln$$(N_{\rm_{ u}}/g_{_{\rm u}})$ errors from the integrated
intensities. The linear least-squares fit (solid line) gives a
rotation temperature of 200 $\pm$ 57 K.} \vspace{3mm}
\end{figure}

\begin{figure}[]
\vspace{0mm}
\includegraphics[angle=270,scale=.72]{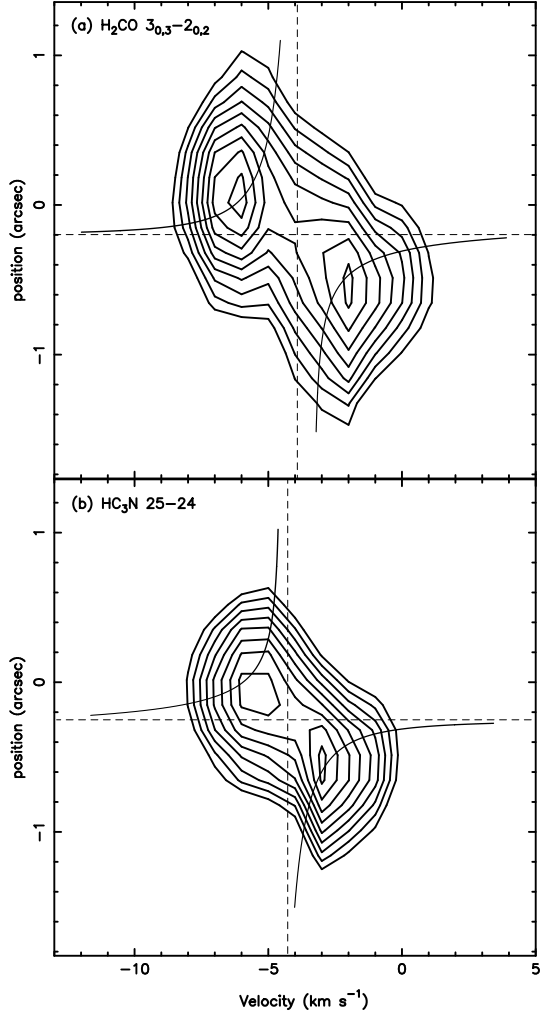}
\vspace{-2mm}\caption{  (a) The position-velocity plot of the $\rm
H_{2}CO$ $3_{_{0,3}}$-$2_{_{0,2}}$ emission at a position angle of
$58^{\circ}$ along the major axis of the disk (Zhang et al. 1999).
The contour levels in observed data are plotted at every $2\sigma$.
$1\sigma$ = 0.6 Jy. Position offsets is ($0.6^{\prime\prime}$,
$1.0^{\prime\prime}$). (b) The position-velocity plot of the
 $\rm HC_{3}N$ $24$-$23$  emission at a position angle of
$58^{\circ}$ along the major axis of the disk. The contour levels in
observed data are plotted at every $2\sigma$. $1\sigma$ = 0.5 Jy.
Position offsets is ($0.6^{\prime\prime}$, $1.2^{\prime\prime}$).
The horizontal dashed line marks the center of the disk, and the
vertical dashed line marks the cloud systemic velocity. The full
lines show Keplerian rotation curve.}
\end{figure}

\end{document}